\documentclass[11pt,twoside]{article}
\usepackage{asp2010}

\def\arcmin{\hbox{$^\prime$}}
\def\arcsec{\hbox{$^{\prime\prime}$}}

\def\OIII{[O\,\textsc{iii}]}
\def\SIII{[S\,\textsc{iii}]}
\def\HII{H\,\textsc{ii}}

\resetcounters

\begin{document}

\title{Planetary Nebula Surveys: Past, Present and Future}

\author{Quentin A. Parker$^{1,2}$ and David J. Frew$^1$
\affil{$^1$Department of Physics and Astronomy, Macquarie University, NSW 2109, Australia}
\affil{$^2$Australian Astronomical Observatory, PO Box 296, Epping, NSW 1710, Australia}
}

\begin{abstract}
In this review we cover the detection, identification and astrophysical importance of planetary nebulae (PN).  The legacy of the historic Perek \& Kohoutek and Acker et al. catalogues is briefly covered before highlighting the more recent but significant progress in PN discoveries in our Galaxy and the Magellanic Clouds. We place particular emphasis on the major MASH and the IPHAS catalogues, which, over the last decade alone, have essentially doubled Galactic and LMC PN numbers.  We then discuss the increasing role and importance that multi-wavelength data is playing in both the detection of candidate PN and the elimination of PN mimics that have seriously biased previous PN compilations. The prospects for future surveys and current efforts and prospects for PN detections in external galaxies are briefly discussed due  to their value both as cosmic distance indicators and as kinematical probes of galaxies and dark matter properties.\\
\noindent{\bf Keywords.}\hspace{12pt}Stars: post-AGB -- planetary nebulae: general -- surveys: general
\end{abstract}

\section{Introduction: the astrophysical value of planetary nebulae}
Planetary nebulae (PN) are amongst the most photogenic and complex of celestial phenomena but are also amongst the most important to properly understand. This is because their brief flowering provides a unique window into the late-stage evolution of low- to intermediate-mass stars. They are important probes of nucleosynthesis processes, mass-loss physics and Galactic abundance gradients, and, because their progenitor stars dominate all stars above one solar mass, PN are responsible for a large fraction of the chemical enrichment of the interstellar medium, including the seeding of pre-biotic carbon between the stars. 

Furthermore, their rich emission-line spectra enable detection to large distances.  These emission lines allow the determination and analysis of chemical abundances and permit the estimation of shell expansion velocities and ages, and so probe the physics and timescales of stellar mass loss (e.g. Iben 1995).  The  measured radial velocities can trace the kinematic properties of  observed PN,  enabling us to decide if they belong to a relatively young or old stellar population. The kinematic properties of PN in galaxy halos also give strong constraints both on the mass distributions and formation processes of giant elliptical galaxies (e.g. Bekki \& Peng 2006; Douglas et al. 2007), making them useful kinematical probes for understanding the structure of galaxies, and to test whether a galaxy contains a substantial amount of dark matter (e.g. Romanowsky et al. 2003; Herrmann \& Ciardullo 2009). 

The PN formation rate also gives the death rate of stars born billions of years ago. They thus directly probe Galactic stellar and chemical evolution (Maciel \& Costa 2003). Their beautiful, complex morphologies provide clues to their formation, evolution, mass-loss processes, and the shaping role that may be played  by magnetic fields, binary central stars (e.g. Moe \& De Marco 2006; Frew \& Parker 2007; De Marco, Hillwig \& Smith 2008; De Marco 2009; Miszalski et al. 2009a,b) or massive planets (e.g. Soker  \& Subag 2005). As the central star fades to become a white dwarf and the nebula expands, the integrated flux, surface brightness and radius change in ways that can be predicted by current stellar and hydrodynamic theory (e.g. Perinotto et al. 2004). 

Importantly,  the ensemble PN luminosity function (PNLF; Ciardullo 2010) in a given galaxy is sufficiently well behaved that it can act as a powerful distance calibrator, or standard candle, to determine the scale of the Universe to better than 10\% (e.g. Ciardullo et al. 2002; Feldmeier, Jacoby \& Phillips 2007), but how and why it works so well is not properly understood. Unravelling the detailed form of the PNLF in resolvable populations of different metallicity in the Galactic Bulge (Kovacevic et al. 2010), LMC (Reid \& Parker 2010)  and the local Galactic disk (Frew 2008) is helping to address this issue. In all these ways PN are  powerful astrophysical tools making them valuable targets for discovery in our own Galaxy, the Local Group, and beyond. 

\section{The first PN discoveries 1764--1999: from Herschel to Acker and Kohoutek}
The first known observation of a PN, the now famous ``Dumbbell'' nebula or M~27,  was undertaken by  Charles Messier in 1764.  By 1800, a further 33 PN had been added, primarily by William Herschel, and thereafter the class was incrementally added to in a largely ad-hoc manner over the next two centuries. The seminal catalogue of Perek \& Kohoutek (1967), followed by the ESO PN catalogues of Acker et al. (1992, 1996) and the essentially equivalent catalogue of Kohoutek  (2001), compiled these heterogeneous samples of principally optical discoveries into catalogues of between $\sim$1000 and $\sim$1900 ''true'' and candidate PN,  assembled from all sources of discovery going back to Messier. They include the bright PN found in the NGC and IC catalogues, plus major samples from Minkowski (1946, 1947), Haro (1952), Abell (1966), Wray (1966), Henize (1967), Longmore (1977) and Lauberts (1982), amongst others, found either from objective-prism surveys or Schmidt telescope direct broad-band imaging photographic surveys in $B$ (and $R$). The $B$-band includes not only the strong H$\beta$ line 4861\AA~but the key \OIII\ 5007\AA~emission line in the filter wing which is  the strongest optical line in unreddened PN spectra, resulting in the bulk of the early discoveries. 

These discovery papers have been supplemented by numerous investigations over the period between 1980 and 1999 that uncovered small samples, or even individual PN (e.g. Dengel, Hartl \& Weinberger 1980; Cappellaro et al. 1994; Kraan-Korteweg et al. 1996).  These lists were supplemented by small but targeted narrow-band surveys at longer wavelengths such as H$\alpha$ (Beaulieu, Dopita \& Freeman 1999) which has the advantage of  partly alleviating the effects of dust when searching in high-value discovery zones such as the Galactic Bulge.  The scientific legacy of the Perek \& Kohoutek and Acker et al. catalogues is reflected in the 650 and 500 citations each has currently received in the literature, powerfully demonstrating the utility that the provision of consolidated object catalogues can have for facilitating investigations by others.  We also note the parallel discoveries of pre-PN and post-AGB stars over the same period, especially following the IRAS mission, as summarised by Szczerba et al. (2007).

\section{A new golden age of PN discoveries: 2000--2010}
Modest PN discovery programmes continued after the release of the updated Kohoutek catalogue such as the \OIII\ CCD surveys of Boumis et al. (2003, 2006) and the \SIII\ survey of Jacoby \& Van de Steene (2004).  This latter survey uncovered 94 candidate PN, with identifications assisted via follow-up 6~cm radio and 
H$\alpha$ observations. The painstaking work of scrutinising the extant wide-field Schmidt surveys for faint PN has also continued until quite recently, though with diminishing returns (e.g. Kerber et al. 2000, and references therein), assisted via the serendipitous PN discoveries of Whiting et al. (2002, 2007) in searches for dwarf galaxies. Such careful scrutiny of wide-field legacy Schmidt surveys has enjoyed a more recent renaissance through the work of the so-called Deep Sky Hunters (Jacoby et al. 2010) where a valuable new sample of $\sim$100 new PN has been uncovered. The online availability of survey imaging data has enabled teams of amateur astronomers to combine efforts to search for very low-surface brightness PN at higher galactic latitudes not covered by the recent H$\alpha$ surveys. Miszalski et al. (in preparation) is also applying semi-automated search techniques to these legacy surveys with some modest success via the so-called Extremely Turquoise Halo Objects Survey (ETHOS), while Gomez et al. (2010) are utilising SDSS data to look for halo PN. These are all valuable additions.

However, it is the recent advent of powerful, high-resolution, narrow-band H$\alpha$ surveys that has led to a new golden age of PN discovery that is still ongoing. To date, most of these new PN discoveries have been made from the SuperCOSMOS  H$\alpha$ Survey (SHS) covering 4000 square degrees of the Southern Galactic Plane and Magellanic Clouds (Parker et al. 2005),  undertaken on the Australian Astronomical Observatory's UK Schmidt telescope. This survey opened up significant new discovery space for both compact and extended low-surface brightness PN thanks to a spatial resolution  of $\sim$1\arcsec\ and a sensitivity to ionized hydrogen of $\sim$2--5 Rayleighs.  The discoveries are described in the Macquarie/AAO/Strasbourg H$\alpha$ catalogues of Parker et al. (2006a; MASH-I) and Miszalski et al. (2008; MASH-II) which list $\sim$900 and $\sim$350 spectroscopically confirmed Galactic PN respectively. Equivalent  discoveries of $\sim$500 PN in the LMC are described by Reid \& Parker (2006a,b).  The success of the SHS directly inspired the similar INT Photometric H$\alpha$ Survey in the northern hemisphere (IPHAS; Drew et al. 2005) which is also providing a rich seam of ongoing PN discoveries  (e.g. Mampaso et al. 2006; Sabin 2008; Viironen et al. 2009a,b; Sabin et al. 2010), so that the total number of Galactic PN is now nearly 3000 (Frew \& Parker 2010a).

The MASH catalogues alone represent the culmination of a 10 year programme of survey searches, candidate identification, and subsequent confirmatory spectroscopy from 350 nights won on 2-, 4- and 8-m optical telescopes, supplemented by radio and space telescope data.  Note that the PN identification criteria described by Frew \& Parker (2010a,b) were widely applied during generation of the MASH catalogues.  Ongoing refinement of these data sets continues such that overall contamination in MASH by non-PN is low. Furthermore, both MASH and now IPHAS PN catalogues contain samples that are generally more evolved, of lower surface brightness, and are more obscured than those listed in the previous compilations. Consequently, a more representative sample of the true, underlying Galactic PN population across a broader range of evolutionary state is provided. Important samples of local ($<$2~kpc) but extremely low-surface brightness evolved PN have been identified (e.g. Pierce et al. 2004; Frew, Madsen \& Parker 2006; Frew 2008) while the true PN nature of other faint nebulae has been revealed (e.g. Frew, Parker \& Russeil 2006). 

Furthermore, the Perek \& Kohoutek and Acker et al. compilations comprise many PN samples from data with widely varying sensitivity, selection technique, detection efficiency, and spatial coverage. Our inability to fully understand the complex selection effects of these heterogeneous samples renders them problematic for generating a reliable PN luminosity function, for proper kinematical modelling, or for the estimation of key parameters necessary to understand their detailed evolution.  Such problems are greatly reduced in  the MASH and IPHAS samples, which provide the largest, least biased and most homogeneous samples of PN currently available in our Galaxy and the LMC. It is essential that any statistical study of the properties of Galactic PN (e.g. Stanghellini \& Haywood 2010) incorporate MASH and IPHAS PN.

\begin{figure}
\begin{center}
\includegraphics[scale=0.442]{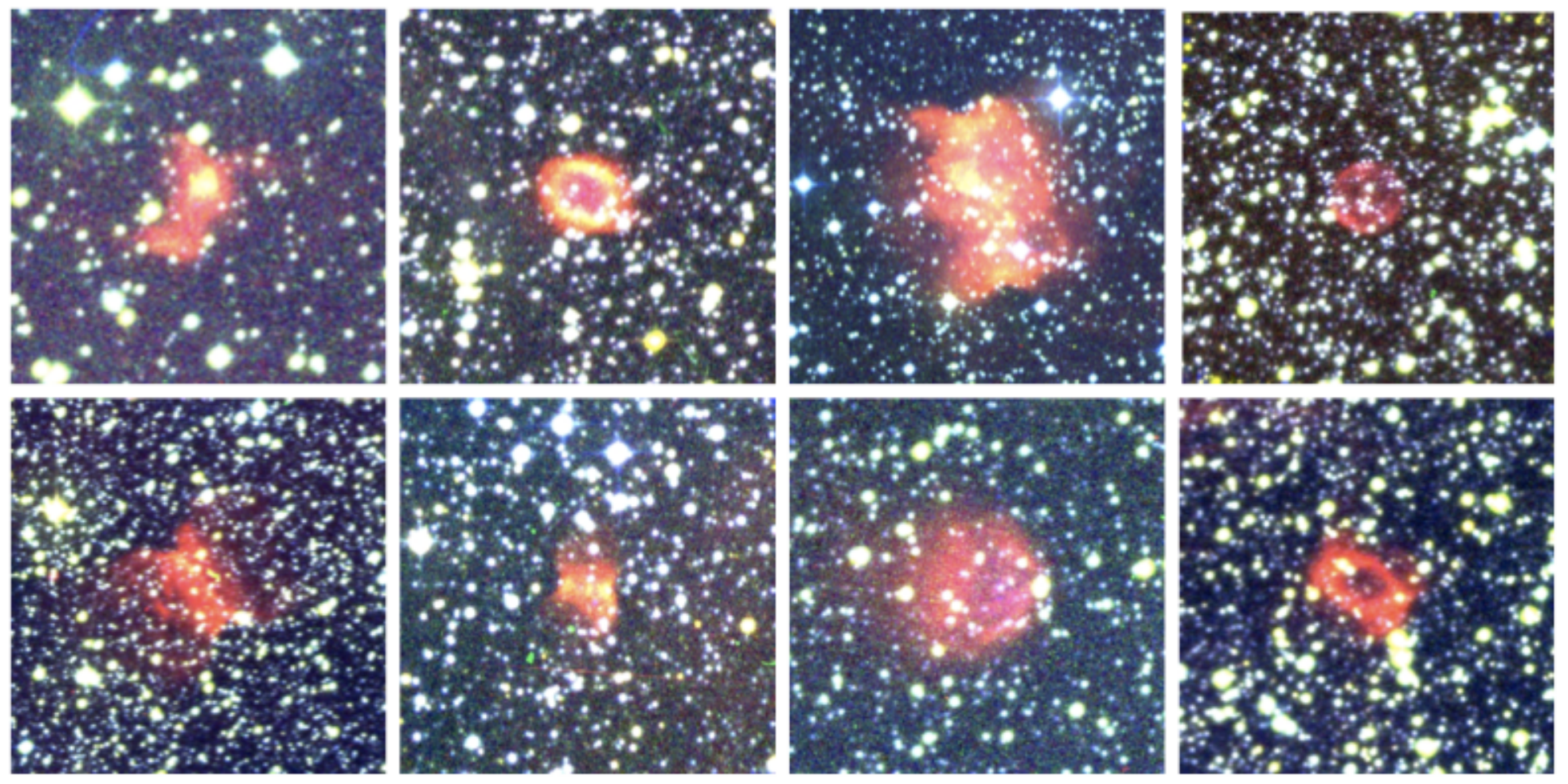}
\caption{A montage of selected MASH-I PN presented as a combined SuperCOSMOS data colour composite. Red: H$\alpha$, Green: SR and Blue: B$_{J}$}
\label{mash-montage}
\end{center}

\end{figure}\subsection{The importance of pure and representative PN samples}
A key point to appreciate is that the utility of PN as astrophysical tools, as elucidated earlier, is predicated on correct identification (e.g. Acker \& Stenholm 1990; Zijlstra, Pottasch \& Bignell 1990). As we show in detail in Frew \& Parker (2010a, and 2010b, these proceedings) it has become increasingly clear that many object types can mimic PN. The availability of wide-field imaging surveys of high resolution and sensitivity across different wavelength regimes in the optical, MIR and radio in particular, not only provide new discovery media but also give enhanced opportunities to explore the broader multi-wavelength properties of PN and their mimics than hitherto possible (e.g. Parker et al. 2006a; Cohen et al. 2007). This has enabled more powerful diagnostic tools to be developed to discriminate between bona-fide PN and interlopers. Again, the reader is referred to the recent review by Frew \& Parker (2010a) where such mimics and the improved tools to identify them are discussed in depth. 

We are now unearthing significant numbers of contaminants in previously existing catalogues. This somewhat undermines the integrity and scientific value of studies based on these previous heterogeneous compilations. As an example, Cohen et al. (2010) use MIR data from GLIMPSE to explore previously known and MASH PN at low Galactic latitude where the selection criteria of Frew \& Parker (2010a) were broadened to incorporate GLIMPSE  data. This led to only a 5\% fraction of MASH PNe being eliminated as contaminants, compared to $\sim$45\% culled from the previously known PNe across the same survey region. This illustrates the uneven nature of the pre-MASH catalogues whose entries were not always subjected to robust diagnostic techniques. It is extremely important that these new selection criteria are brought to bear on the pre-MASH catalogues, to avoid biasing the interpretations and analyses performed on these data. For example, Kwok et al. (2008) base a detailed MIR analysis on 30 PN, taking their PN identifications as given. However, application of our new discriminatory tools reveal 16 to be \HII\ region contaminants (Cohen et al. 2010). Furthermore, in the construction of a carefully evaluated sample of PN within a nearby 1.0~kpc local volume, Frew (2008) found that a significant number of objects, perhaps up to 20\%, are, in fact Str\"omgren spheres around hot pre-white dwarf or subdwarf stars.  These stars are simply ionizing the surrounding ambient ISM through which they are passing; see Frew \& Parker (2006), Madsen et al. (2006), and Frew et al. (2010) for specific examples.

An important goal is the determination of the total PN population in our Galaxy. Estimates of this number have a direct bearing on our understanding of stellar evolution theory, Galactic chemical enrichment rates, and Galactic ecology.  Important related questions concern the formation mechanisms of  PN.  For example, is a common envelope origin (Moe \& De Marco 2006, De Marco 2009, and these proceedings) a pre-requisite for generating most PN?  Are different evolutionary pathways to PN-like nebulae possible (Frew \& Parker 2010a)?  Although the current number of Galactic PN is only around 3000, this is double what it was just a decade ago as a result of the MASH surveys in the southern hemisphere (Parker et al. 2006a; Miszalski et al. 2008), and now from the equivalent IPHAS survey in the north (Sabin et al. 2010;  Viironen et al. 2009a, b). However, this is still far short of our best estimates of the expected Galactic PN population based on population synthesis models or extrapolations from local space densities (see Jacoby et al. 2010). We still urgently need to improve the detection completeness for Galactic PN. Progress is being made (e.g. refer to the previous review of Parker et al. 2006b) and with new results and surveys coming on stream, the diversity of pathways to PN formation may be finally revealed.

\section{The power of multi-wavelength observations:  refining existing samples and uncovering hidden populations}
The large majority of Galactic and Magellanic Cloud PNe have been historically discovered in the optical domain, first with broadband imaging and objective prism techniques, then using \OIII\ narrow-band imaging. More recently, as illustrated by the large numbers of  MASH and IPHAS discoveries, deep H$\alpha$ narrow-band imaging has been used to isolate what is normally the second most intense optical emission line in unobscured PNe. The success of these surveys is due to the ability to find PN in more dusty regions where this line can be much stronger.
 
Jacoby \& Van de Steene (2004) were the first to conduct searches for PN in obscured regions  using narrow-band filters to isolate lines further to the NIR to be less affected by dust.  Large numbers  of candidate PN have also been selected via  IRAS mid-infrared colours (e.g. Preite-Martinez 1988; Pottasch et al. 1988; Ratag et al. 1990; Garc\'ia-Lario et al. 1997)  but success rates for confirming these as PN via follow-up spectroscopy have been quite modest to date (e.g. S\'uarez et al. 2006). 
The availability of large-scale, wide-field Galactic surveys  at high angular resolution in several optical and near/mid-infrared and radio passbands (e.g. SHS, IPHAS,  SDSS, 2MASS, UKIDSS, GLIMPSE, MIPSGAL, MSX, AKARI, MOST, NVSS) provides unprecedented opportunities to combine multi-wavelength detections to refine selection techniques to effectively eliminate contaminants (Frew \& Parker 2010a). 

\subsection{The value of mid-IR surveys}
Mid-IR imagery allows the detection of extremely reddened PN invisible at optical wavelengths (e.g. Cohen et al. 2005; Phillips \& Ramos-Larios 2008). Such new data allow us to investigate quantitative differences in PN multi-wavelength characteristics (e.g. Cohen et al. 2007, 2010) and  relate these to the PN age, type and chemistry. Accounting for the non-ionised component of PN via MIR data also improves the derivation of mass-loss estimates and other key physical parameters,  helping us to better understand the role of PN in the chemical evolution of our Galaxy. Recently, Carey et al. (2009) and Mizuno et al. (2010) have noted over 400 compact ($<$1\arcmin) ring, shell, and disk-shaped sources in the Galactic plane at 24$\mu$m in {\em Spitzer} MIPSGAL images.  We believe that many of these will turn out to be strongly reddened, high-excitation PN with only a minority being circumstellar nebulae around massive stars (cf. Gvaramadze, Kniazev \& Fabrika 2010; Wachter et al. 2010).  PN can be strong mid-IR sources because of PAH emission, fine-structure line emission, and molecular and thermal dust emission within the PN shells, and in any circumnuclear disks. Cohen et al. (2010) analysed 136 optically detected PN and candidates from the GLIMPSE-I survey with the goal of developing robust, multi-wavelength classification criteria to augment existing diagnostics and provide pure PN samples.
The ultimate goal is to recognise PN using only MIR and radio characteristics. This will then enable us to trawl for PN effectively even in highly obscured regions of the Galaxy that are impossible to access using traditional techniques. We expect to uncover a rich population of hidden PN.

\section{The future of PN surveys and key science projects} 
Despite the recent significant progress in PN surveys much work still remains to be done in realising the full astrophysical potential of this fascinating phenomenon across different galaxies and environments. This can only be done in an unbiased way when truly representative samples are available. Many outstanding problems remain, such as resolving the role and importance of binary central stars and establishing clear links between central star properties and their PN --- currently only $\sim$25\% of PN central stars have been unequivocally identified, while the proportion with spectra is even less (e.g. Weidmann \& Gamen 2010).  Additionally, the importance and evolutionary role of the [WR] class remains elusive --- see De Pew et al. (2010, and these proceedings). Furthermore, only $\sim$150 Galactic PN have accurate abundances, while the origin and development of the complex morphologies and asymmetries, on both large and small physical scales, is poorly known. A more complete inventory of Galactic PN is urgently needed as our current best estimates of the Galactic PN population range from 6000 to 30,000, still way in excess of those actually discovered, although the known LMC sample is now getting close to the best population estimates (see Reid \& Parker 2006b).
Immediate progress will come from the essentially complete IPHAS H$\alpha$ survey where large numbers of PN candidates are now being  followed-up spectroscopically.  Confirmatory spectroscopy is also needed for the newly found DSH, SDSS, ETHOS and other candidate objects.  However, infrared spectroscopy will be needed for most of the new GLIMPSE and MIPSGAL discoveries which have no optical counterparts, though some of the less reddened examples can be identified in the optical (e.g. Fesen \& Milisavljevic 2010). The honing of MIR identification techniques (e.g. Cohen et al. 2010) offers prospects for finding PN in external galaxies at greater distances with the James Webb Space Telescope.
Finally, future narrow-band and multi-waveband surveys of the Galactic plane (VPHAS+, VVV, etc.) and out-of-plane regions (e.g. SkyMapper) will provide excellent additional discovery capability.  The future is bright.

\acknowledgements 
QAP thanks the SOC for an Invited Review, and Macquarie University and the Australian Astronomical Observatory for  travel funding.
\section*{References}
{\small
Acker, A., et al. 1992,  Strasbourg-ESO Catalogue of Galactic Planetary Nebulae (Garching)\\
Acker, A., Marcout J. \& Ochsenbein F. 1996,  First Supplement to the SECPGN (Strasbourg)\\ 
Acker, A. \& Stenholm, B. 1990, A\&AS, 86, 219\\
Abell, G.O. 1966, ApJ, 144, 259\\ 
Beaulieu, S.F., Dopita, M.A. \& Freeman, K.C. 1999, ApJ, 515, 610\\
Bekki, K. \& Peng, E.W. 2006, MNRAS, 370, 1737\\
Boumis, P., Paleologou, E.V., Mavromatakis, F. \& Papamastorakis, J. 2003, MNRAS, 339, 735\\
Boumis, P., et al. 2006, MNRAS, 367, 1551\\ 
Cappellaro, E., Sabbadin, F., Salvadori, L., Turatto, M. \& Zanin, C. 1994, MNRAS, 267, 871\\
Carey, S.J., et al. 2009, PASP, 121, 76\\
Ciardullo, R., et al. 2002, ApJ, 577, 31\\
Ciardullo, R. 2010, PASA, 27, 149\\ 
Cohen, M.C., et al. 2005, ApJ, 627, 446\\
Cohen, M.C., et al. 2007, ApJ, 669, 343\\ 
Cohen, M.C., et al. 2010, MNRAS, submitted\\ 
De Pew, K., et al.  2010, MNRAS, submitted\\
De Marco, O. 2009, PASP, 121, 316\\
De Marco, O., Hillwig, T.C. \& Smith, A.J. 2008, AJ, 136, 323\\
Dengel, J., Hartl, H. \& Weinberger, R. 1980, A\&A, 85, 356\\
Douglas, N.G., et al. 2007, ApJ, 664, 257\\
Drew, J.E., et al. 2005, MNRAS, 362, 753\\ 
Feldmeier, J.J., Jacoby, G.H. \& Phillips, M.M. 2007, ApJ, 657, 76\\
Fesen, R.A. \& Milisavljevic, D. 2010, AJ, 139, 2595\\
Frew, D.J. 2008, PhD thesis, Macquarie University\\
Frew, D.J. \& Parker, Q.A. 2006, IAUS, 234, 49\\
Frew, D.J. \& Parker, Q.A. 2007, APN4 Proceedings (IAC Elec. Pub), p. 475\\
Frew, D.J. \& Parker, Q.A. 2010a, PASA, 27, 129\\ 
Frew, D.J. \& Parker, Q.A. 2010b, APN5 Proceedings (Ebrary), arXiv:1010.5003\\
Frew, D.J., Madsen, G.J. \& Parker, Q.A. 2006, IAUS, 234, 395\\
Frew, D.J., Parker, Q.A. \& Russeil, D. 2006, MNRAS, 372, 1081\\ 
Frew, D.J., Madsen, G.J., O'Toole, S.J., \& Parker Q.A. 2010, PASA, 27, 203\\ 
Garc\'ia-Lario, P., Manchado, A., Pych, W. \& Pottasch, S.R. 1997, A\&AS, 126, 479\\
Gomez, T., et al. 2010, BAAS, 42, 472\\
Gvaramadze, V.V., Kniazev, A.Y.  \& Fabrika, S., 2010, MNRAS, 405, 1047\\
Haro, G. 1952, Bolet\'in de los Observatorios de Tonantzintla y Tacubaya,  1, 1\\
Henize, K.G. 1967, ApJS, 14, 125\\ 
Herrmann, K.A. \& Ciardullo, R. 2009, ApJ, 705, 1686 \\
Iben, I., Jr. 1995, Phys. Reports, 250, 2\\
Jacoby, G.H. \& Van de Steene, G. 2004, A\&A, 419, 563\\
Jacoby, G.H. et al. 2010, PASA, 27, 156\\ 
Kerber, F., Furlan, E., Roth, M., Galaz, G. \& Chanam\'e, J.C. 2000, PASP, 112, 542\\
Kohoutek, L. 2001, A\&A, 378, 843\\ 
Kovacevic, A. et al. 2010, MNRAS, submitted\\
Kraan-Korteweg, R., et al. 1996, A\&A, 315, 549\\ 
Kwok, S., Zhang, Y., Koning, N., Huang, H.-H. \& Churchwell, E. 2008, ApJS, 174, 426\\ 
Lauberts, A. 1982, The ESO/Uppsala Survey of the ESO (B) Atlas (Garching: ESO)\\
Longmore, A.J. 1977, MNRAS, 178, 251\\ 
Maciel, W.J. \& Costa, R.D.D. 2003,  IAUS, 209, 551\\
Madsen, G.J., Frew, D.J., Parker, Q.A., Reynolds, R.J. \& Haffner L.M. 2006, IAUS, 234, 455\\
Mampaso, A., et al. 2006, A\&A, 458, 203\\
Minkowski, R. 1946, PASP, 58, 305\\ 
Minkowski, R. 1947, PASP, 59, 257\\
Miszalski, B., et al. 2008, MNRAS, 384, 525 (MASH-II)\\ 
Miszalski, B., Acker, A., Moffat, A.F.J., Parker, Q.A. \& Udalski, A. 2009a, A\&A, 496, 813\\ 
Miszalski, B., Acker, A., Moffat, A.F.J., \& Parker, Q.A. 2009b, A\&A, 505, 249\\
Mizuno, D.R., et al. 2010, AJ, 139, 1542\\
Moe, M. \& De Marco, O., 2006, ApJ, 650, 916\\ 
Parker, Q.A. et al. 2005, MNRAS, 362, 689\\ 
Parker, Q.A. et al. 2006a, MNRAS, 373, 79 (MASH-I)\\ 
Parker, Q.A., Acker, A., Frew, D.J. \& Reid, W.A. 2006b, IAUS, 234, 1\\ 
Perek, L. \& Kohoutek, L. 1967, Catalogue of Galactic Planetary Nebulae (Prague)\\ 
Perinotto, M., Sch\"onberner, D., Steffen, M. \& Calonaci, C. 2004, A\&A, 414, 993\\
Phillips, J.P. \& Ramos-Larios, G. 2008, MNRAS, 386, 995\\
Pierce, M.J., Frew, D.J., Parker Q.A. \& K\"oppen, J. 2004, PASA, 21, 334\\
Pottasch, S.R., Olling, R., Bignell, C. \& Zijlstra, A.A. 1988, A\&A, 205, 248\\
Preite-Martinez, A. 1988, A\&AS, 76, 317\\
Ratag, M.A., Pottasch, S. R., Zijlstra, A.A. \& Menzies, J. 1990, A\&A, 233, 181\\
Reid, W.A. \& Parker, Q.A. 2006a, MNRAS, 365, 401\\
Reid, W.A. \& Parker, Q.A. 2006b, MNRAS, 373, 521\\
Reid, W.A. \& Parker, Q.A. 2010, MNRAS, 405, 1349\\
Romanowsky, A.J., et al. 2003, Science, 301, 1696\\  
Sabin, L. 2008, PhD thesis, University of Manchester\\ 
Sabin, L. et al. 2010, PASA, 27, 166\\ 
Soker, N., \& Subag, E. 2005, AJ, 130, 2717\\
Stanghellini, L. \& Haywood, M. 2010, ApJ, 714, 1096\\
Su\'arez, O., et al. 2006, A\&A, 458, 173\\ 
Szczerba, R., Si\'odmiak, N., Stasi\'nska, G. \& Borkowski, J. 2007, A\&A, 469, 799\\ 
Viironen, K. et al. 2009a, A\&A, 502, 113\\
Viironen, K. et al. 2009b, A\&A, 504, 291\\
Wachter, S., et al. 2010, AJ, 139, 2330\\ 
Weidmann, W.A. \& Gamen, R. 2010, A\&A, in press, arXiv:1010.5376 \\ 
Whiting, A.B., Hau, G.K.T. \& Irwin, M. 2002, ApJS, 141, 123\\
Whiting, A.B., Hau, G.K.T., Irwin, M. \& Verdugo, M. 2007, AJ, 133, 715\\
Wray, J.D. 1966, PhD thesis, Northwestern University\\ 
Zijlstra, A., Pottasch, S. \&  Bignell, C., 1990, A\&AS, 82, 273
}



\end{document}